\magnification=1200
\hsize=15truecm
\vsize=23truecm
\baselineskip 17 truept
\voffset=-0.5truecm
\parindent=0cm
\overfullrule=0pt

\def\pmb#1{\leavevmode\setbox0=\hbox{$#1$}\kern-.025em\copy0\kern-\wd0
\kern-.05em\copy0\kern-\wd0\kern-.025em\raise.0433em\box0}

\def\Ai{\hbox{\hbox{${\cal A}$}}\kern-1.9mm{\hbox{${/}$}}}
\def\Vi{\hbox{\hbox{${\cal V}$}}\kern-1.9mm{\hbox{${/}$}}}
\def\Di{\hbox{\hbox{${\cal D}$}}\kern-1.9mm{\hbox{${/}$}}}
\def\lam{\hbox{\hbox{${\lambda}$}}\kern-1.6mm{\hbox{${/}$}}}
\def\D{\hbox{\hbox{${D}$}}\kern-1.9mm{\hbox{${/}$}}}
\def\A{\hbox{\hbox{${A}$}}\kern-1.8mm{\hbox{${/}$}}}
\def\V{\hbox{\hbox{${V}$}}\kern-1.9mm{\hbox{${/}$}}}
\def\parz{\hbox{\hbox{${\partial}$}}\kern-1.7mm{\hbox{${/}$}}}
\def\B{\hbox{\hbox{${B}$}}\kern-1.7mm{\hbox{${/}$}}}
\def\R{\hbox{\hbox{${R}$}}\kern-1.7mm{\hbox{${/}$}}}
\def\si{\hbox{\hbox{${\xi}$}}\kern-1.7mm{\hbox{${/}$}}}

\catcode`@=11
%
%---------------------------- \lsim \gsim ----------------------------------
%
%    Simboli di minore o circa uguale, maggiore o circa uguale.
%
\def\lsim{\mathchoice
  {\mathrel{\lower.8ex\hbox{$\displaystyle\buildrel<\over\sim$}}}
  {\mathrel{\lower.8ex\hbox{$\textstyle\buildrel<\over\sim$}}}
  {\mathrel{\lower.8ex\hbox{$\scriptstyle\buildrel<\over\sim$}}}
  {\mathrel{\lower.8ex\hbox{$\scriptscriptstyle\buildrel<\over\sim$}}} }
\def\gsim{\mathchoice
  {\mathrel{\lower.8ex\hbox{$\displaystyle\buildrel>\over\sim$}}}
  {\mathrel{\lower.8ex\hbox{$\textstyle\buildrel>\over\sim$}}}
  {\mathrel{\lower.8ex\hbox{$\scriptstyle\buildrel>\over\sim$}}}
 {\mathrel{\lower.8ex\hbox{$\scriptscriptstyle\buildrel>\over\sim$}}} }

\def\gsu{\raise4pt\hbox{\kern5pt\hbox{$\sim$}}\lower1pt\hbox{\kern-8pt
\hbox{$>$}}~}
\def\lsu{\raise4pt\hbox{\kern5pt\hbox{$\sim$}}\lower1pt\hbox{\kern-8pt
\hbox{$<$}}~}

\def\croce{\displaystyle / \kern-0.2truecm\hbox{$\backslash$}}
\def\lqua{\lower4pt\hbox{\kern5pt\hbox{$\sim$}}\raise1pt
\hbox{\kern-8pt\hbox{$<$}}~}
\def\gqua{\lower4pt\hbox{\kern5pt\hbox{$\sim$}}\raise1pt
\hbox{\kern-8pt\hbox{$>$}}~}
\def\mma{\lower1pt\hbox{\kern5pt\hbox{$\scriptstyle <$}}\raise2pt
\hbox{\kern-7pt\hbox{$\scriptstyle >$}}~}
\def\mmb{\lower1pt\hbox{\kern5pt\hbox{$\scriptstyle >$}}\raise2pt
\hbox{\kern-7pt\hbox{$\scriptstyle <$}}~}
\def\mmc{\lower4pt\hbox{\kern5pt\hbox{$<$}}\raise1pt
\hbox{\kern-8pt\hbox{$>$}}~}
\def\mmd{\lower4pt\hbox{\kern5pt\hbox{$>$}}\raise1pt
\hbox{\kern-8pt\hbox{$<$}}~}
\def\croce{\displaystyle / \kern-0.2truecm\hbox{$\backslash$}}
%
%
%---------------------------  \quadratello ---------------------------------
%
%
\def\quad@rato#1#2{{\vcenter{\vbox{
        \hrule height#2pt
        \hbox{\vrule width#2pt height#1pt \kern#1pt \vrule width#2pt}
        \hrule height#2pt} }}}
\def\quadratello{\mathchoice
\quad@rato5{.5}\quad@rato5{.5}\quad@rato{3.5}{.35}\quad@rato{2.5}{.25} }
%
%------------------------ caratteri grassetto speciali -----------
%
\font\s@=cmss10\font\s@b=cmbx8
\def\reali{{\hbox{\s@ l\kern-.5mm R}}}
\def\m{{\hbox{\s@ l\kern-.5mm M}}}
\def\k{{\hbox{\s@ l\kern-.5mm K}}}
\def\naturali{{\hbox{\s@ l\kern-.5mm N}}}
\def\interi{{\mathchoice
 {\hbox{\s@ Z\kern-1.5mm Z}}
 {\hbox{\s@ Z\kern-1.5mm Z}}
 {\hbox{{\s@b Z\kern-1.2mm Z}}}
 {\hbox{{\s@b Z\kern-1.2mm Z}}}  }}
\def\complessi{{\hbox{\s@ C\kern-1.7mm\raise.6mm\hbox{\s@b l}\kern.8mm}}}
\def\toro{{\hbox{\s@ T\kern-1.9mm T}}}
\def\unity{{\hbox{\s@ 1\kern-.8mm l}}}
%
%------------------------- bold math. it. --------------------------
%
\font\bold@mit=cmmib10
\def\setbmit{\textfont1=\bold@mit}
\def\bmit#1{\hbox{\textfont1=\bold@mit$#1$}}
%
%-----------------------------------------------------------------
\catcode`@=12

\null
\rightline{DFPD 00/TH/52}

\vskip 2truecm

\centerline{\bf AN ORDER PARAMETER RECONCILING ABELIAN AND}

\centerline{\bf CENTER DOMINANCE IN SU(2) YANG--MILLS THEORY}

\vskip 1truecm

\centerline{J. Fr\"ohlich$^1$, P.A. Marchetti$^2$}
\vskip 0.5truecm

\centerline{\sl 1 -- Theoretical Physics, ETH--H\"onggerberg, CH--8093,
Z\"urich}

\centerline{\sl 2 -- Dipartimento di Fisica, Universit\`a di Padova and
INFN -- }

\centerline{\sl Sezione di Padova, I -- 35131 Padova}

\vskip 2truecm
\midinsert
\baselineskip 12pt
\leftskip 1.3truecm
\rightskip 1.3truecm
\parindent 0truecm

\centerline{\bf Abstract}
\vskip 1.5truecm
\quad
We analyze previously proposed order parameters for the 
confinement - deconfinement transition in lattice SU(2) 
Yang--Mills theory, 
defined as vacuum expectation value (v.e.v.) of monopole fields in abelian 
projection gauges. We show that they exhibit some inconsistency in the 
treatment of small scales, due to a violation of Dirac quantization 
condition for fluxes.

We propose a new order parameter avoiding this inconsistency. It can be 
interpreted as v.e.v. of the field of a regular monopole in any abelian 
projection gauge, but it is independent of the choice of the abelian 
projection. Furthermore, being constructed in terms of surfaces of center 
vortices, it has also a natural interpretation in the approach of center 
dominance.
\endinsert

\vskip 0.5truecm

{\bf 1. Introduction}
\vskip 0.3truecm

\quad
There now appears to be a general consensus about the idea that color 
confinement in Yang Mills theories is caused by the condensation of some 
topological defects. However it is still debated whether the important 
defects are center vortices or magnetic monopoles.

\quad
The first proposal was brought forward, in the '70, by 't Hooft, Polyakov, 
Mack, Nielsen and Olesen [1] 
and others and has received renewed interest in recent times, 
stimulated by numerical simulations [2]; presently it is usually named 
``center dominance".

\quad
The basic idea is drawn from an analogy with ferromagnets and may be 
roughly phrased as follows. The expectation value of a temporal Wilson loop 
in four dimensions
can be seen, by fixing the temporal gauge, as a product of two--point spin 
correlation functions of three dimensional 
non linear $\sigma$ models coupled to gauge fields, one for each time 
involved in the Wilson loop.
If the non--linear $\sigma$--models are disordered, their spin correlation 
functions have
an exponential decay. This implies an area law for the Wilson loop of the
four--dimensional gauge theory and 
hence, according to Wilson's criterion, confinement. (Actually, a rigorous 
proof of a mathematically precise version of this idea has been established
in [3]). The defects responsible for disorder in the non linear $\sigma$ 
models  are 
vortices. For $SU(N)$ theories they are ${\bf Z}_N$ center 
vortices. 

\quad
The alternative with monopoles as the relevant defects [4] has been put on a 
concrete basis by 't Hooft [5], who suggested to make explicit the monopoles
in $SU(N)$ Yang--Mills theories,
by performing a gauge fixing that leaves a maximal Cartan torus,
$U(1)^{N-1}$,
unbroken. These gauges are called abelian projection gauges. The resulting
abelian gauge theories can be 
rewritten in terms of ``photons", gauge fields of a theory with 
gauge group the decompactification of the residual gauge group, and
monopoles.
The monopoles 
corresponding to this decompactification are those expected to condense in 
the confinement phase.
\quad
 
To define in $SU(2)$ Yang-Mills theory, to which this paper is addressed 
explicitly, an abelian projection gauge, 't Hooft 
suggested to construct some
scalar field $X$ taking values in $su (2)$, as a function of the gauge 
connection and transforming in 
the adjoint representation of the gauge group, $SU(2)$. To perform the 
gauge fixing one imposes the constraint that $X$ is diagonal.

\quad
The diagonal component of the $SU(2)$--connection in this gauge plays the 
role of the ``photon field", the off--diagonal components are charged with 
respect to the residual gauge group $U(1)$.

\quad
The points in space--time where two eigenvalues of the matrix $X$ 
coincide are the positions of the monopoles in this gauge. Confinement is 
then believed to emerge as a consequence of monopole condensation in the 
form of a ``dual--Meissner" effect.

\quad
Together with the assumption that the effect of the charged off--diagonal 
degrees of freedom are qualitatively irrelevant for the description of the 
low--energy physics, the above scenario is usually called ``abelian 
dominance".

Two natural questions arise in the ``abelian dominance" scenario:

1) Is there a ``monopole field operator" which plays the r\^ole of an 
order parameter for the confinement--deconfinement transition, i.e. with 
vanishing expectation value in the deconfined phase and non--vanishing 
expectation value in the confined phase, so that the transition can be 
interpreted as due to a ``monopole condensation"?

2) does the choice of the field $X$ influence the behaviour of the 
expectation value of the monopole field operator?

In [6,7] an attempt has been made to give a positive answer to the first 
question on the basis of a circle of ideas which can be presented as 
follows.

\quad
In electrodynamics with charged scalar field $\phi$ one can construct 
gauge--invariant charged field operators and their correlation functions 
adapting the Dirac recipe [8], dressing the local non--gauge invariant field 
$\phi (x)$ with a cloud of soft photons, represented by  multiplication by a
phase factor with argument given by the gauge field $\vec A$ weighted by a
classical Coulomb field $\vec E(x)$.

\quad
In abelian gauge theories there is a natural notion of duality exchanging 
the r\^ole of charges and monopoles. One can obtain monopole correlation 
functions from gauge--invariant charged correlation functions by a duality 
transformation; in particular this applies to monopoles in $U(1)$ lattice 
gauge theory.

\quad
In [6,7] it was suggested to apply this construction to the $SU(2)$ gauge 
theory in the abelian projection gauge and it was shown [7] how to render this 
construction gauge--invariant.

\quad
Starting from the expectation value of the monopole operator constructed in
this manner, Montecarlo simulations show that

1) this monopole field operator indeed behaves as a good order parameter 
signaling the confinement -- deconfinement transition;

2) the (physical) temperature of the transition is independent of the choice
of the scalar field $X$ [9]. 

\quad
This approach however  
presents a foundational problem in spite of its great numerical success: it 
is inconsistent with Dirac's quantization condition of fluxes. This, in 
turn, implies an inconsistency of the treatment of small scales.

\quad
To understand the origin of the Dirac quantization condition, let us
consider 
the electrodynamics of electrically and magnetically charged point--like 
particles.

The equations of motion proposed by Dirac read:

$$
\partial^\mu F_{\mu\nu} (x) = q_e j^e_\nu (x), \ \epsilon^{\mu\nu\rho\sigma}
\partial_\nu F_{\rho\sigma} (x) = q_m j^m_\nu (x) \eqno(1.1)
$$

where  $q_e j^e_\nu$ and $q_m j_\nu$ are the electric and magnetic currents
generated by the particles, $q_e$ and $q_m$ being their electric and magnetic
charges.

Since equation (1.1) implies current conservation

$$
\partial^\nu j^e_\nu = 0 = \partial^\nu j^m_\nu,
$$

Poincar\'e's lemma ensures  the existence of antisymmetric tensor 
fields $n^e_{\rho\sigma}, n^m_{\rho\sigma},$ such that

$$
j^e_\mu = \epsilon_{\mu\nu\rho\sigma} \partial_\nu n^e_{\rho\sigma}, \ 
j^m_\mu =\epsilon_{\mu\nu\rho\sigma} \partial_\nu n^m_{\rho\sigma}. 
\eqno(1.2)
$$

However these fields are determined only modulo the transformation

$$
n^\#_{\rho\sigma} \rightarrow n^\#_{\rho\sigma} + \partial_{[\rho} 
\lambda ^\#_{\sigma]} \eqno(1.3)
$$

where $\#= e,m$ and $\lambda^\#_\sigma$ is a vector field.

Schwinger proposed an action leading to the Dirac equations (1.1):

$$
S(A, j^e, n^m) = \int \Bigl[{1\over 2} (\partial_{[\mu} A_{\nu]} + q_m 
n^m_{\mu\nu})^2 (x) + q_e (A_\mu j^{e\mu}) (x)\Bigr] d^4 ,
$$

where if the support of the current $j^m_\mu$ lies on a curve $\gamma$, 
then $n^m_{\mu\nu}$ should be taken as a surface current whose support is a
surface $\Sigma$, with boundary $\gamma$. The current $n^m$ describes the 
Dirac strings attached to the magnetic monopoles whose worldlines are given
by $\gamma$.

\quad
The quantum theory corresponding to the classical action $S(A, j^e, n^m)$ 
is well defined, provided the partition function 

$$
\int {\cal D} A e^{iS (A, j^e, n^m)}
$$

is independent of the choice of the surface $\Sigma$, with $n^m$ satisfying
(1.2), i.e. invariant under the gauge transformations (1.3).

\quad
If one chooses $n^m_{\rho\sigma}$ and $n^{\prime m}_{\rho\sigma}$ corresponding 
to the surfaces $\Sigma$ and $\Sigma^\prime$, then the parameter 
$\lambda_\sigma^m$
appearing in (1.3) is dual to a volume--current with support in the volume $V$ 
whose boundary is given by the closed surface difference of $\Sigma^\prime$ and 
$\Sigma$.

The consistency condition turns out to be given by

$$
q_m  q_e \int \lambda^m_\mu j^{e \mu} d^4 x \in 2\pi {\bf Z}. \eqno(1.4)
$$

Since $j_\mu$ are line currents with integral coefficients one can 
recognize in (1.4) the Dirac quantization condition:

$$
q_e q_m \in 2\pi {\bf Z}. \eqno(1.5)
$$

\quad
From equation (1.4) it follows as consistency requirement that the 
integral over an 
arbitrary volume of any 
electric current appearing in the partition function,  
multiplied by ${q_m\over 2\pi}$ must be an integer. 
One can easily prove that this condition extends to all physical 
correlation functions.

\quad
However in the construction of the gauge invariant correlation functions 
for charged fields following the Dirac recipe, one introduces the electric 
smooth current $E$, whose integral over a generic volume $V$ is a real 
number, thus violating Dirac quantization condition. One may say that in 
the presence of the electric current $E$ the position of the Dirac strings 
(of monopoles) crossing the support of $E$ become visible, thus introducing
a physical inconsistency.

\quad
The monopole correlation functions may be obtained from the charged 
correlation functions by duality, therefore in the presence of dynamical 
charges 
their construction following Dirac's recipe encounter the same inconsistency.
This extends also to the constructions of monopoles for $SU(2)$ presented in 
[6,7], 
since it involves charged degrees of freedom corresponding to the 
off--diagonal components of the Yang--Mills gauge field.

\quad
Nevertheless, since in $U(1)$ gauge theories the constraint of Dirac 
quantization is expected to become irrelevant at large distances,
i.e. in the scaling limit, one might still imagine that the low--energy 
physics, (e.g. the deconfinement transition 
temperature),
should not be affected by the above problems.

\quad
In this paper we propose a new order parameter for the 
confinement -- deconfinement transition, defined in terms of the expectation 
value of a (regular) monopole field operator, which avoids the 
inconsistency discussed above.

Our construction is a variant of that proposed for the 't Hooft--Polyakov 
monopoles in the Georgi--Glashow model [10] and, as a basic difference with 
respect to previous constructions, the correlation functions of monopoles 
are obtained in terms of sheets of center vortices.

The new order parameter exhibits the following features:

1) it respects the Dirac quantization condition

2) it is naturally independent of the choice of a $U(1)$ subgroup of 
$SU(2)$, needed in the abelian projection

3) if one chooses an abelian projection gauge we argue that in the scaling 
limit it approaches the order parameter constructed in [7] in 
correspondence to that projection gauge

4) it creates a bridge between ``abelian" and ``center dominance" 
suggesting how one can reconcile the two approaches.

\quad
Although our discussion is heuristic and partly conjectural, the overall 
picture that emerges is consistent with known mathematical estimates
and numerical lattice simulations.

In order to make the paper selfcontained, we start by recalling in section 2
Dirac recipe and its dual, sketching the corresponding construction of 
charged and monopole Green function in the euclidean formalism.

\quad
In section 3 we review the modification of Dirac's ansatz proposed 
in order to satisfy the Dirac quantization condition if 
dynamical charges and monopoles coexist.

In sect. 4 we sketch how one defines the magnetic charge in $SU(2)$ theory.

In sect. 5 we review, in the euclidean formalism, earlier constructions 
of monopole field in $SU(2)$, based on the dual of Dirac's recipe.

In sect. 6 we define our new monopole field and the corresponding order 
parameter and discuss the link to previous constructions.

In sect. 7 we outline the connection between ``abelian" and ``center 
dominance" suggested by our  order parameter.

In order to simplify our formulas in the rest of the paper we use the 
language of forms and currents, both in the continuum and on the lattice.
The basic definitions of this formalism are reviewed in an appendix.

\vskip 0.5truecm

{\bf 2. Dirac's ansatz and its dual}
\vskip 0.3truecm

\quad
We start by discussing a simple model, scalar QED, where only 
electric dynamical charges appear, but no magnetic monopoles. We show 
how to construct gauge--invariant, charged field operators 
following Dirac's ansatz.

\quad
Let $\phi$ be a massive scalar field with charge $e$ coupled to an abelian
gauge field, $A_\mu$, described in terms of a real 1--form $A$, with 
classical action

$$
S(A, \phi) = \int\Bigl[{1\over 2} (\partial_{[\mu} A_{\nu]})^2 (x) 
+{1\over 2} |\partial_\mu -  ie A_\mu) \phi |^2 (x) + 
{m^2\over 2} \bar\phi \phi (x)\Bigr] d^4 x
$$

or, in the notation of differential forms,

$$
S(A, \phi) = {1\over 2} || dA||^2 + {1\over 2} || (d - ie A) 
\phi ||^2 + {m^2\over 2} ||\phi||^2. \eqno(2.1)
$$

Dirac's ansatz  can be formulated as follows: Let $\hat\phi$ 
and $\hat A$ denote the quantum field operators corresponding to the 
classical fields $\phi$ and $A$  and let $E^{\vec x}_\mu dx^\mu \equiv 
E^{\vec x}$
denote the 1--form corresponding to the classical electromagnetic field 
generated by a pointlike unit charge located at $\vec x$ in ${\bf R}^3$. 
Then Dirac's charged field is defined, heuristically, by the formula:

$$
\hat\phi (E^{\vec x}) = \hat\phi (\vec x) e^{i\int_{{\bf R}^3} \hat A 
\wedge^* E^{\vec x}}. \eqno(2.2)
$$

This construction has been rendered rigorous (in the presence of 
an ultraviolet regulation)
in [11], within the indefinite metric approach (Gupta--Bleuler
gauge).

\quad
There is however an alternative route to construct charged fields. 
One starts from euclidean Green functions and then invokes the 
Osterwalder--Schrader reconstruction theorem [12].
In approximate terms it works as follows: One 
constructs gauge--invariant euclidean correlation functions for charged 
fields obeying the O.S. axioms (essentially translation invariance, 
reflection (O.S.) positivity and clustering) from which one  can  
reconstruct a Hilbert space  of physical  states containing the vacuum 
vector $\Omega$, a unitary representation of space--time translations, 
whose  generators satisfy the spectral condition, and which leaves
$\Omega$ 
invariant, and quantum field operators.

\quad
There is also a version of the reconstruction theorem that applies 
to lattice theories.  

\quad
A euclidean version of Dirac's ansatz is then obtained by replacing the 
quantum field (2.2) by a euclidean field

$$
\phi(E^x) = \phi (x) e^{i\int_{{\bf R}^4} A \wedge^* E^x}= \phi (x)
e^{i(A, E^x)}  \eqno(2.3)
$$

where $E^x$ is the 1--current in ${\bf R}^4$ given by 

$$
E^x (y) = E^{\vec x} (\vec y) \delta (y^0 - x^0). \eqno(2.4)
$$

\quad
With a lattice regularisation euclidean correlation functions of these 
fields, $\langle \prod_i \phi (E^x_j) \prod_j \bar\phi (E^y_j) \rangle$,
have been 
proved to satisfy the O.S. axioms [13]. Here $\langle \cdot \rangle$ 
denotes the 
euclidean expectation value corresponding to the action (2.1) and 
$\bar\phi(E^x)$ is the complex conjugate of $\phi (E^x)$.

\quad
It is useful for later purposes to notice that one can obtain a 
representation of the correlation functions of $\phi (E)$ as partition 
functions of a gas of closed electric currents coupled to $A$, by 
integrating out $\phi$.

\quad
As an example consider the two-point function

$$
\langle \phi (E^x) \bar\phi (E^y)  \rangle = {1\over Z} \int DA e^{-{1\over 
2} || dA||^2} {\rm det} (-\Delta_{eA} + m^2)
$$
$$
e^{ie (A,(E^x - E^y))} \langle \phi(x) \bar\phi (y) \rangle (A) 
\eqno(2.5)
$$

where $\langle \cdot \rangle (A)$ denotes the (normalised) expectation value 
corresponding to the action of $\phi$ coupled to $A$, viewed as an ``external" 
field, and $\Delta_{eA}$ is the covariant Laplacian.

\quad
Using a euclidean version of Feynman's 
path--integral formula for the quantum--mechanical time evolution kernel 
$(e^{s \Delta_{eA}}) (x,y)$, one obtains formally 

$$
\langle \phi (x) \bar\phi (y) (A) \rangle = (-\Delta_{eA} + m^2)^{-1}(x,y)
= \int^\infty_0 ds e^{-sm^2} (e^{s \Delta_{eA}}) (x,y)=
$$
$$
= \int^\infty_0 ds e^{-sm^2} \int\limits_{\scriptstyle 
q(0)=x \atop\scriptstyle q(s)=y} 
{\cal D} 
q (t) e^{-\int^s_0 
\Bigl[{1\over 4}\dot q^2 (t) + e A_\mu (q(t)) \dot q^\mu (t) \Bigr] dt}.
$$

\quad
If with every trajectory $\{q_\mu (t), t\in [0,s]\}$, $s\in
{\bf R}_+$, we associate the 1--current

$$
j_{xy} (y) = \int dt \dot q_\mu (t) \delta (q (t)-y) dy^\mu,
$$

$(^*j_{xy}$ is  Poincar\'e dual of the trajectory, see appendix), then,
one finds

$$
\langle \phi (x) \bar\phi (y) \rangle (A) = \int {\cal D}\mu (j_{xy}) 
e^{ie(A,j_{xy})},
$$

for a suitable measure ${\cal D}\mu (j_{xy})$.

\quad
One can also express the determinant in (2.5) in terms of a sum 
over closed current networks {\bf j}, so that,
for a suitable measure ${\cal D}\mu ({\bf j})$ on the current networks 
{\bf j}, one obtains

$$
\langle \phi(E_x) \bar\phi (E_y) \rangle = \int{\cal D} A e^{-{1\over 2} 
||dA||^2} \int{\cal D}\mu ({\bf j}) {\cal D}\mu (j_{xy}) e^{i e (A, 
(E^x - E^y + j_{xy} + {\bf j}))}
$$ 
$$
\bigl[\int{\cal D} A e^{-{1\over 2} ||dA||^2} \int {\cal D}\mu ({\bf j}) 
e^{i e (A, {\bf j})} \bigr]^{-1}. \eqno(2.6)
$$

We notice that, while the current networks {\bf j} in the 
partition function (the denominator of (2.6)) are all integer--valued, the
currents appearing in the numerator of (2.6) also involve a real--valued 
term $(E^x - E^y)$.

We now sketch how one can obtain monopole correlation functions in a dual 
theory. For $U(1)$--gauge theories in $d$=4 dimensions, S--duality is a 
transformation 
mapping the correlation functions of the original gauge theory onto those
of a dual 
gauge theory, exchanging the role of charges and monopoles. The underlying 
idea can be presented as follows.

Let $S(dA)$ denote a gauge--invariant action for the gauge field $A$, 
written in terms of its curvature 2--form $dA$. Introducing an auxiliary 
2--form $F$ we can write the partition function as

$$
Z = \int {\cal D} A e^{-S(dA)} = \int {\cal D} F e^{-S(F)} \delta (dF), 
\eqno(2.7)
$$

since the solution of the constraint

$$
dF=0 \eqno(2.8)
$$

is given by $F= dA$ and the Jacobian is field--independent.

We now express the constraint in (2.8) by a Fourier representation of the
$\delta$ functional:

$$
\delta (dF) = \int {\cal D} \tilde A e^{i\int F\wedge d\tilde A} \eqno(2.9)
$$

where $\tilde A$ is a new gauge field, the ``dual of $A$". We define the 
dual action $\tilde S (d\tilde A)$ through the functional integral Fourier 
transform

$$
e^{-\tilde S (d\tilde A)} \equiv \int {\cal D} F e^{-S(F)} e^{i\int F\wedge 
d\tilde A}.
\eqno(2.10)
$$

Plugging (2.9) and (2.10) in to (2.7) we obtain 

$$
Z = \int {\cal D} F e^{-S(F)} \int {\cal D} \tilde A e^{i\int F \wedge 
d\tilde A} = \int {\cal D} \tilde A e^{-\tilde S (d\tilde A)} 
$$

where the last term gives the partition function of the dual theory.

The same procedure proves that duality exchanges the Wilson loop,
which can be related to worldlines of a charged 
particle--antiparticle pair, with the Wegner -- 't Hooft disorder 
operator [14], related to worldlines of a monopole--antimonopole pair.

Let $\Sigma$ be a 2--dimensional surface;  the 
Poincar\'e--dual current is also denoted by $\Sigma$. The Wilson 
loop $W_\alpha (\Sigma)$,
$\alpha \in {\bf R}$, is defined in terms of the 2--form $F$ appearing in 
(2.7) as

$$
W_\alpha (\Sigma) = e^{i \alpha\int_\Sigma dA} = e^{i\alpha \int_\Sigma F} =
e^{i\alpha \int F\wedge \Sigma}. \eqno(2.11)
$$

In the same model the Wegner--'t Hooft disorder field is given by 

$$
D_\alpha (\Sigma) = e^{-[S(F + \alpha \Sigma) - S (F)]}. \eqno(2.12)
$$

The duality transformation acts on such fields as follows:

$$
\langle W_\alpha (\Sigma) \rangle = \int{\cal D}  F e^{-S(F)} \int{\cal D} 
\tilde A e^{i\int F\wedge d\tilde A} e^{i\alpha \int F\wedge \Sigma} = \int 
{\cal D} \tilde A e^{-\tilde S (d\tilde A + \alpha \Sigma)}
$$
$$
= \int {\cal D} \tilde F  e^{-\tilde S (\tilde F + \alpha \Sigma)} \int{\cal 
D} A e^{i\int \tilde F \wedge dA} = \langle D_\alpha (\Sigma) \rangle^\sim
\eqno(2.13)
$$

where $\langle \cdot \rangle^\sim$ denotes the expectation value in the 
dual theory.

To apply duality to scalar QED, we notice 
that the representation of the partition function in terms of the current 
networks appearing in (2.6) 
can be viewed as the expectation value of a weighted sum of Wilson loops, 
$e^{i e\int_{\Sigma({\bf j})} dA}$, with a weighting measure 
${\cal D}\mu ({\bf j})$ in a gauge theory with action

$$
S(dA) = {1\over 2} ||dA||^2,
$$

if we associate to every current configuration 
{\bf j} a 2--current $\Sigma ({\bf j})$ satisfying

$$
d\Sigma ({\bf j}) = ^*{\bf j}. \eqno(2.14)
$$

As a result the partition function of the dual theory can 
be written as 

$$
\tilde Z = \int {\cal D}\mu ({\bf j}) \int {\cal D} \tilde A e^{-{1\over 2} 
||d\tilde A + 
e \Sigma ({\bf j})||^2}. \eqno(2.15)
$$

Obviously it corresponds to a Maxwell theory with gauge 
potential $\tilde A$ coupled to monopoles, whose worldlines are described 
by {\bf j}; $\Sigma({\bf j})$ can then be identified as the surface spanned 
by the Dirac strings of the monopoles.

The partition function is independent of the choice of the Dirac strings, 
since a different choice $\Sigma^\prime ({\bf j})$ also satisfying

$$
d\Sigma^\prime ({\bf j}) = ^*{\bf j},
$$

differs from $\Sigma ({\bf j})$ by an exact 2--form 
$dV$ which can be absorbed by a change of variables $\tilde A \rightarrow 
\tilde A + V$; (Poincar\'e's lemma). 

By performing the shift 

$$
A \rightarrow A + e \delta \Delta^{-1} \Sigma ({\bf j})
$$

and using the Hodge decomposition for $\Sigma ({\bf j})$ (see equation 
(A.4) in the appendix) one can alternatively rewrite the 
partition function (2.15) as

$$
\tilde Z = \int {\cal D}\mu ({\bf j}) \int {\cal D} \tilde A e^{-{1\over 2} || 
d\tilde A + e \delta \Delta^{-1}  {}^*{\bf j} ||^2}. \eqno(2.16)
$$

The term $\delta\Delta^{-1} {}^*{\bf j}$ can  be interpreted as the magnetic
field generated by the magnetic current networks {\bf j}.
One can obtain the two--point monopole correlation function, $\langle m (B^x)
\bar m (B^y)\rangle^\sim$, from (2.6) by applying the duality transformation. 

Setting $B= ^* E,$ and $e= \tilde g$ (the magnetic charge in the 
dual theory), one finds

$$
\langle m (B^x) \bar m (B^y) \rangle^\sim = \langle \phi (E^x) \bar\phi 
(E^y)\rangle = 
$$
$$
{1\over \tilde Z} \int{\cal D} \mu ({\bf j}) {\cal D}\mu 
(j_{xy}) {\cal D} \tilde A e^{-{1\over 2} ||d\tilde A + \tilde g 
\delta \Delta^{-1} 
(^*{\bf j} + ^*j_{xy} + B^x - B^y)||^2}. \eqno(2.17)
$$

The magnetic 3--current $B^x(y)$ can be related to the magnetic field 
strength (2--form) $B^{\vec x}$ of a classical monopole located at $\vec x$ 
in ${\bf R}^3$ by

$$
B^x (y) = B^{\vec x} (\vec y) \wedge \delta (x^0 - y^0) dy^0. \eqno(2.18)
$$

These ideas can be applied to general models of gauge theories with only 
dynamical charges or monopoles, and they can be rendered mathematically 
rigorous for lattice theories. 

As an example of a lattice theory with monopoles, one may consider the $U(1)$ 
gauge theory in the Villain or Wilson formulation.

The basic field is a $U(1)$ lattice gauge field $\theta$, i.e. a 
$U(1)$--valued 1--form, and in the Villain formulation one must introduce 
a $2\pi {\bf Z}$--valued two--form field $n$.
The actions are given by

$$
S_V (\theta, n) = \beta ||d\theta + n ||^2
$$
$$
S_W (\theta) = \beta \sum_p (1- \cos (d\theta)_p) \eqno(2.19)
$$

where the subscript $V$ stands for ``Villain" and $W$ for ``Wilson". 
The monopole two--point functions are given by

$$
G_V (x,y) = {1\over Z_V} \int \prod_{<xy>} d\theta_{<xy>} \sum_n 
e^{-\beta ||d\theta + n + 2\pi \delta \Delta^{-1} (\omega+ B_x - B_y)||^2}
$$
$$
G_W (x,y) = {1\over Z_W} \int \prod_{<xy>} d\theta_{<xy>} e^{-\beta\sum_p 
(1-\cos (d\theta + 2\pi \delta \Delta^{-1} (\omega + B_x - 
B_y)_p))}, \eqno(2.20)
$$

where $Z$ denotes the partition function of the model, 
the summation in the Villain model is over  
the $2\pi {\bf Z}$--valued $n$--configurations and $\omega$ is a 
3--form Poincar\'e dual of a path joining $\{x \}$ to $\{ y\}$.

The Green function for monopoles in the Villain model can be recast in a 
form similar to that appearing in eq. (2.17) by defining a real--valued 
1--form $A$ and a $2\pi{\bf Z}$--valued 3--form $m$, with

$$
A = \theta + \delta \Delta^{-1} n, \qquad m= dn.
$$

Then

$$
G_V (x,y)= {1\over Z_V} \sum_{m:dm=0} \int \prod_{<xy>} dA_{<xy>}
$$
$$
e^{-\beta||dA + \delta \Delta^{-1} (m+ 2\pi (B_x+B_y + \omega))||^2}. 
\eqno(2.21)
$$

Gauge--fixing for $A$ (or quotienting w.r.t. gauge transformations) is 
understood in the $A$--measure in (2.21).

Finally, by defining , e.g. in the Wilson formulation, a disorder field

$$
D_\omega (B^x , B^y) = e^{-[S_W (d\theta+ 2\pi \delta \Delta^{-1} 
(B^x+\omega-B^y)) - S_W (d\theta)]}, \eqno(2.22)
$$

one can express the monopole two--point function as an expectation value of
the disorder field:

$$
G_W (x,y) = \langle D_\omega (B^x, B^y)\rangle.
$$

It has been rigorously shown in [13,15] (see also [16]) 
that this disorder field is a good 
order parameter for the confinement--deconfinement transition in $d=4 \ U(1)$
gauge theories, $G (x,y)$ approaches a 
finite value in the confining phase and vanishes in the deconfined phase,
as $|x-y| \rightarrow\infty$.

One can the interpret the non--vanishing asymptotic value of $G(x,y)$ as a 
signal of monopole--condensation.

\vskip 0.3truecm
{\bf 3. A modified Dirac ansatz consistent with Dirac's quantization 
condition}
\vskip 0.3truecm

\quad 
In this section we discuss the modification of the construction of the 
previous section needed when dynamical charges and monopoles coexist. We 
start by showing that, as it stands, the above construction becomes 
inconsistent in this enlarged setting.

\quad
Consider, for example, a model of ``compact" scalar QED with a scalar field
$\phi$ of electric charge $e$ and monopoles of magnetic charge $g$, whose 
partition function is given by

$$
Z = \int {\cal D} \tilde\mu ({\bf j}) \int {\cal D} A 
e^{-{\beta \over 2} || dA + g\Sigma ({\bf j})||^2} \quad
\int {\cal D} \phi {\cal D} \bar\phi e^{-\int{1\over 2} \bar\phi 
(-\Delta_{eA} + m^2)\phi}, \eqno(3.1)
$$

where $\Sigma ({\bf j})$ are the {\bf Z}--valued Dirac strings of the 
monopoles, and the measure ${\cal D} \tilde\mu ({\bf j})$ is derived from
the action of a matter field coupled to $A$, e.g. a complex scalar field 
$\tilde\phi$ of mass $\tilde m$, through an equation like 

$$
{\int {\cal D} \tilde{\bar\phi}{\cal D} \tilde\phi e^{-{1\over 2} \int
\tilde{\bar\phi}
(-\Delta_{eA} +\tilde m^2)\tilde\phi} \over \int {\cal D}
\tilde{\bar\phi} {\cal D}
\tilde\phi e^{-{1\over 2} \int \tilde{\bar\phi} (-\Delta +
m^2)\tilde\phi}}
= \int {\cal D}\tilde\mu ({\bf j}) e^{ie({\bf j}, A)} \eqno(3.2)
$$

\quad
In order for $Z$ to be physically well defined, it should be 
independent of the choice of the Dirac strings satisfying $^*d\Sigma({\bf 
j}) = {\bf j}$.

\quad
It is convenient to rewrite the integral over $\phi$ and $\bar\phi$ in 
terms of {\bf Z}--valued electric current networks ${\bf l}$, as

$$
{\int {\cal D} \phi {\cal D} \bar\phi e^{-{1\over 2} \int \bar\phi
(-\Delta_{eA} + \tilde m^2)\phi} \over \int{\cal D} \phi {\cal D} \bar\phi 
e^{-{1\over 2} \int \bar\phi (-\Delta + m^2)\phi}} = \int{\cal D} \mu 
({\bf \l}) e^{i
e (A, {\bf l})}. \eqno(3.3)
$$

\quad
Then the situation analysed in the introduction emerges, and the
consistency
condition that guarantees that Dirac strings are invisible, in the sense
that $\Sigma ({\bf j}) \rightarrow 
\Sigma^\prime ({\bf j})$ is a symmetry of the 
theory, becomes invariance of $Z$ under the shift 
$A\rightarrow A + g V ({\bf j})$, for $V({\bf j})$ a {\bf Z}--valued 
1--current satisfying $dV ({\bf j}) = \Sigma^\prime ({\bf j})-\Sigma ({\bf
j})$, i.e.
$$
e^{i e g (V({\bf j}), {\bf l})} = 1. \eqno(3.4)
$$

The condition (3.4) 
is the Dirac quantization condition at the level of
currents, and it is satisfied, provided the Dirac quantization condition 
for charges

$$
e g \in 2\pi {\bf Z} \eqno(3.5)
$$

holds.

\quad
If one tries to construct the 2--point function of the charged field 
according to Dirac's ansatz, as in the previous section, one meets an 
inconsistency, since, in  contrast to (3.4),

$$
e^{i e g (V({\bf j}), ({\bf l} + l_{xy} + E^x - E^y))} \not = 1,
\eqno(3.6)
$$

even if (3.5) holds.

The origin of this problem is a violation of Dirac's quantization condition
at the level of currents, due to the introduction of the real--valued 
current $E$.

\quad
Let us first consider this problem for minimal charges:$ e g=2\pi$. One 
might envisage avoiding the difficulty encountered above by replacing the 
electric current  $E^x$ by an electric ``Mandelstam string" [17]
$\gamma^x$
carrying a unit flux along a path starting from $x$ and reaching infinity, 
at fixed time.

\quad
Such a current would still satisfy $\delta \gamma^x = \delta_x,$
and it does not violate Dirac's quantization condition. However, the 
current $\gamma^x$ does not decay at infinity (in contrast to $E^x$) and, 
as a consequence, infrared divergences appear in the construction of charged 
correlation functions based on this ansatz.

In fact, lattice calculations [13] suggest that an abelian gauge theory 
with massive monopoles and charges scales to a gaussian gauge theory at 
large distances, in the Coulomb phase.

Hence, to every Mandelstam string $\gamma^x$ is associated an infinite 
positive self--energy, $\sim (\gamma^x, \Delta^{-1} \gamma^x)$, and the 
interaction energy between two Mandelstam strings $\gamma^x, \gamma^y$ of 
opposite charge is infinite and negative, $\sim - (\gamma^x, \Delta^{-1} 
\gamma^y)$, because the strings have infinite length.

\quad
Even if the selfenergies are subtructed off, via a multiplicative 
renormalisation, the interaction between the two strings cannot be removed 
without violating reflection positivity, because it depends on the
distance 
between $x$ and $y$. A violation of reflection positivity would, 
however, render impossible the reconstruction of charged quantum
fields. 

\quad
A possible way  to circumvent this infrared divergence was suggested in 
[10] : one has to replace a fixed Mandelstam string $\gamma^x$ by a sum
over 
fluctuating Mandelstam strings weighted by a measure ${\cal D} \nu 
(\gamma^x)$ supported on strings fluctuating so strongly that their 
interaction energy remains finite even in the limit of infinite length.

The strings appearing in the construction of the correlation functions of 
charged fields should then converge to a common point at infinity.

As will be reviewed later, it has been shown in [10] that there exists a 
lattice regularised complex measure ${\cal D}\nu_E (\gamma^x)$ on {\bf
Z}--valued 
1--currents $\gamma^x$ satisfying

$$
\delta \gamma^x = \delta_x
\eqno(3.7)
$$

such that

i) the correlation functions for the euclidean fields

$$
\phi (x) \int {\cal D} \nu_E (\gamma^x) e^{i e(A, \gamma^x)},
$$
$$
\bar\phi (y) \int {\cal D} \nu_E (\gamma^y) e^{-ie (A, \gamma^y)} 
\eqno(3.8)
$$

satisfy a lattice version of the O.S. axioms; and 

ii) in the scaling limit,

$$
\int{\cal D} \nu_E (\gamma^x) e^{i e (\gamma^x, A)} \sim e^{i e (E^x, A)}, 
\eqno(3.9)
$$

up to a multiplicative renormalisation, where $E$ is the electric 
``Coulomb" field. Thus on large scales, the sum over fluctuating 
Mandelstam strings reproduces the behaviour of phase factor appearing in
the Dirac ansatz.
(This has been verified in [10], in a gaussian approximation). 
\quad
Equation 
(3.9) suggests that, at large scales, the measure ${\cal D} \nu_E 
(\gamma^x)$ mimics an approximate $\delta -$ function peaked around
$E^x$.

\quad
Let us suppose that an appropriate variant of the O.S. axioms is satisfied 
by expectation values of the euclidean fields (3.8), (as, follows formally, 
from their definition). Then from their correlation functions one can 
reconstruct quantum field operators

$$
\hat\phi (E^x) ,\quad \hat{\bar\phi} (E^y).
$$

See [10,13] for details.

If we consider compact scalar QED and the Dirac quantization condition
(3.5) is 
satisfied in the more general form $g= {2\pi \over e} q, \quad q \in 
{\bf Z}\quad (q \not = 0)$, then we can repeat the above construction of charged
quantum fields, replacing the {\bf Z}--valued currents $\gamma^x$ with 
{\bf Z}/q--valued currents $\gamma^x$ (satisfying (3.7)).

Let us describe how to construct a monopole--monopole Green function. 

We associate a ${\bf Z}/q$--valued 2--current $\Sigma (\gamma^x - \gamma^y 
+ j_{xy})$ to an integral 1--current $j_{xy}$ satisfying $\delta j_{xy}
= \delta_y - \delta_x $ and a pair of ${\bf Z}/q$--valued currents 
$\gamma^x$ and $\gamma^y$ ( with $\delta\gamma^z= \delta_z$, $z=x,y$),
such that

$$
{}^*d\Sigma (\gamma^x - \gamma^y + j_{xy})= \gamma^x - \gamma^y + 
j_{xy}.\eqno(3.10)
$$

We then define a disorder field by setting 

$$
D(\Sigma(\gamma^x - \gamma^y + j_{xy}))= e^{-{\beta\over 2}\{||dA + 
g(\Sigma(\gamma^x-\gamma^y+j_{xy})+\Sigma ({\bf j}))||^2 - ||dA + 
\Sigma ({\bf j}) ||^2\}} .\eqno(3.11)
$$

One can easily verify that $\langle D\bigl(\Sigma (\gamma^x + \gamma^y + 
j_{xy})\bigr)\rangle$ depends only on $\gamma^x+\gamma^y+ j_{xy}$ and 
\underbar{not} on a specific choice of $\Sigma$, because different choices
can be mapped onto one another by a shift of $A$.

\quad
The monopole 2--point function is given by

$$
\int {\cal D} \nu_E (\gamma^x) {\cal D} \nu_E (\gamma^y) {\cal D} \tilde\mu
(j_{xy}) \langle D(\Sigma (\gamma^x + \gamma^y + j_{xy}))\rangle, 
\eqno(3.12)
$$

where ${\cal D}\tilde\mu (j_{xy})$ is the measure defined through the
equation 

$$
{\int {\cal D} \tilde{\bar\phi} {\cal D} \tilde\phi e^{-{1\over 2} \int
\widetilde{\bar\phi} 
(-\Delta_{eA} + \tilde m^2)\tilde\phi} \tilde{\bar\phi} (y) \tilde\phi
(x)\over 
\int{\cal D} \tilde{\bar\phi} {\cal D} \tilde\phi e^{-{1\over 2}
\int \tilde{\bar\phi} 
(-\Delta_{eA} +\tilde m^2)\tilde\phi}}
= \int {\cal D} \tilde\mu (j_{xy}) e^{ie(j_{xy}, A)}, 
$$

for arbitrary $A$.

Denoting by $\langle \cdot  \rangle^\sim$ the expectation value in the 
dual model, one can easily verify that (3.12) equals

$$
\langle \tilde{\bar\phi} (E^y) \tilde\phi (E^x)\rangle^\sim.
$$

\quad
From correlation functions of disorder fields such as (3.11) one can 
obtain, via O.S. reconstruction, monopole field operators, $\hat m(B^x)$,
where 
$B^x =^*E^x$ encodes the infrared behaviour of the soft photon
cloud accompanying the monopole, (which is an ``infra--particle").

In particular, for $y^0 < 0 < x^0$, the correlation function (3.12) is 
equal to

$$
\langle \hat m (B^y) \Omega| \hat m (B^x) \Omega \rangle.
$$

\quad
The modification of Dirac's recipe and its dual suggested above can be 
adapted to any abelian theory of coexisting charges and monopoles, and it 
can be made precise for lattice theories. The measure ${\cal D}\nu_E$ can
then be 
constructed as follows; (we omit some technical details about b.c., see
[10 ]): We introduce a $U(1)$--valued scalar field $\chi$ of period $2\pi 
q$ defined on a three--dimensional hyperplane $\Lambda_{x^0}$, at fixed time 
$x^0$, minimally coupled to the gauge field $A$. An action is given e.g. by

$$
S(\chi, A)= {\xi\over 2} \sum_{<xy>\in\Lambda_{x^0}} \bigl[1 - \cos
{(d\chi + 
A)_{<xy>}\over q}\bigr], \eqno(3.13)
$$

and we denote the corresponding expectation value by $\langle \cdot 
\rangle_{x^0} (A).$

\quad
Then the two--point function,  $\langle e^{i\chi_x}
e^{-i\chi_y}\rangle_{x^0}
(A)$, can be expressed in terms of ${\bf Z}/q$--valued 1--currents 
$\gamma_{xy}$ satisfying $\delta\gamma_{xy} = \delta_x - \delta_y$ as

$$
\langle e^{i\chi_x} e^{-i\chi_y} \rangle_{x^0} (A) = \int{\cal D} \nu 
(\gamma_{yx}) e^{i(A, \gamma_{xy})}, \eqno(3.14)
$$

where $x^0= y^0$.

\quad
If $\xi$ is sufficiently large (and if the field strength 
$\{(dA)_p\}_{p\in\Lambda_{x^0}}$  does not fluctuate much) the system 
described by $\chi$ is in a phase where the symmetry

$$
\chi \rightarrow \chi + {\rm const}
\eqno(3.15)
$$

is spontaneously broken [3]. The associated Goldstone boson is a real 
field, $\lambda$, describing a spin wave. It corresponds to the 
decompactification of the range of $\chi$. Deviations from the theory 
described by non--interacting spin waves are due to vortices. They are 
believed to be irrelevant at large scales (in the sense of the 
renormalization group), provided the symmetry (3.15) is spontaneously
broken. We 
propose to evaluate (3.14) at large scales, neglecting vortices. The
result is 

$$
{1\over Z} \int {\cal D} \lambda e^{i\lambda_x} e^{-i\lambda_y} 
e^{-{\xi_{ren} \over 2}||d\lambda + A||^2}
$$
$$
= e^{-{1\over 2} ((\delta_x - \delta_y), \Delta^{-1}_{x^0} (\delta_x - 
\delta_y))} e^{i(d\Delta^{-1}_{x_0} (\delta_x - \delta_y), A)}, 
\eqno(3.16)
$$

where $\xi_{ren}$ denotes a renormalised coupling, and $\Delta_{x^0}$ is 
the 3D laplacian on $\Lambda_{x^0}$.

One can easily verify that 

$$
E^x - E^y = d \Delta^{-1}_{x_0} (\delta_x - \delta_y). \eqno(3.17)
$$

Sending $y$ to $\infty$ and eliminating the first factor in (3.16) by a 
multiplicative renormalization, one obtains

$$
\bigl(\langle e^{i\chi_x} e^{-i\chi_\infty} \rangle (A)\bigr)_{ren}
\sim e^{i(E^x, A)}
$$

\quad
This suggests that the desired measure is of the form

$$
{\cal D}\nu_E (\gamma^x) \sim  \lim_{y\to\infty} c_y {\cal D}\nu
(\gamma_{xy}),
$$

where $c_y$ denotes a suitable multiplicative renormalisation constant, that
we expect to be finite, on the basis of a gaussian computation performed in
[10].

\quad
As example of lattice theories with dynamical charges and monopole one may 
consider models where a $U(1)$--gauge field $\theta$ is coupled to a 
$U(1)$--valued scalar matter field $\phi$ of charge $q=1,2,...$; (here we
set the elementary charge equal to unity).

In Wilson's formulation, the action is given by

$$
S ={\sum_p} \beta (1 - \cos (d\theta)_p) +
\kappa \sum_{<ij>} (1- \cos (q \theta + d\phi)_{<ij>} ).\eqno(3.17)
$$

According to the recipe explained above, the monopole two--point Green 
function (for monopoles of unit charge) is given by

$$
{1\over Z} \int \prod_{<ij>} d\theta_{<ij>} \prod_i d\phi_i \int {\cal D}
\nu_E
(\gamma^x) {\cal D} \nu_E (\gamma^y)
$$
$$
e^{-\beta \sum_p \bigl(1 - \cos (d\theta + 2 \pi \Sigma(\gamma^x-\gamma^y
+ 
j_{xy}))_p)}
e^{-\kappa \sum_{<ij>} \bigl(1- \cos (q \theta + 
d\phi)_{<ij>}\bigr)},\eqno(3.18)
$$

where $\Sigma (\gamma^x - \gamma^y + j_{xy})$ is a ${\bf Z}/q$--valued 
2--form satisfying (3.10), and $j_{xy}$ is an integral 1--current with 
support on a path connecting $x$ to $y$ and $\delta j_{xy} = \delta_x - 
\delta_y$.

\quad
The Green function (3.18) is independent of the choice of $j_{xy}$ since a 
change in $j_{xy}$ can be compensated by a shift of $\theta$. Hence, in 
contrast to the continuum models previously discussed, no integration over 
$j_{xy}$ is needed.

\vskip 0.3truecm
{\bf 4. The magnetic charge in $SU(2)$ gauge theory}
\vskip 0.3truecm

\quad
We turn to the $SU(2)$ Yang--Mills theory, identifying its magnetic 
monopoles and defining their magnetic charge.

We first discuss these matters in the continuum, where we denote by $A$ the
$SU(2)$--connection form.

Let us consider a cube $c$ in ${\bf R}^3$; its boundary, $\partial c$ is 
homeomorphic to a 2--sphere. The restriction of $A$ to $\partial c$ can be 
viewed as a connection form $\tilde A \equiv A|_{\partial c}$ of an 
$SO(3)$--bundle over $\partial c$.

\quad
$SO(3)$--bundles over $S^2 \simeq \partial c$ are classified by $\pi_1 
(SO(3))\simeq {\bf Z}_2$. The following relation holds between $\tilde A$ 
and the integer class $n$ mod 2 $\in {\bf Z}/2 {\bf Z} \simeq \pi_1
\bigl(SO(3)\bigr)$ 
classifying the corresponding $SO(3)$ bundle:

$$
e^{i\pi n} = e^{i\arg\sum_{p\in\partial c} Tr \bigl(P e^{i\oint_{\partial
p} A^{(p)} }\bigr)}, \eqno(4.1)
$$

where $A^{(p)}$ is the connection 1--form of an $SU(2)$--bundle over the 
face (``plaquette") $p$ of the cube $c$, obtained by lifting the 
$SO(3)$--bundle over $p$.

\quad
If $e^{i\pi n} \not = 1$ one cannot extend the $SO(3)$--bundle to the 
interior of the cube $c$; this signals the presence of an odd number of 
${\bf Z}_2$--monopoles of $SO(3)$ inside $c$ and we identify $e^{i\pi n}$ as
the ${\bf Z}_2$--charge contained in $c$.

\quad
An abelian projection gauge, defined as in the introduction, selects a 
$U(1)$ residual gauge group. By projecting $\tilde A$ to the Cartan 
subalgebra of $su(2)$ corresponding to the residual gauge group one
obtains
a $U(1)$ connection $a$ on $\partial c$.

$U(1)$--bundles over $S^2 \simeq \partial c$ are classified by $\pi_1 
(U(1)) \simeq {\bf Z}$. The relation between the integer $n$ classifying 
the bundle and the connection $a$ is given by

$$
n= {1\over 2\pi} \sum_{p\in\partial c} \int F (a), \eqno(4.2)
$$

where $F(a)$ is the curvature of $a$.

\quad
We identify the magnetic charge contained inside $c$ in the abelian 
projection gauge with the integer $n$, corresponding to the first Chern 
number of $F (a)$.

\quad
If $a$ is derived from $\tilde A$ by abelian projection, the integer $n$ 
appearing in (4.2) is the same appearing (4.1).

\quad
According to a general theorem [18], the classical monopoles associated to 
a $U(1)$ subgroup $SO(3)$ are regular if they carry even magnetic charge.
For these monopoles there are gauge choices for which no singularity of 
Yang--Mills curvature occurs at the monopole position.
Monopoles with odd magnetic charge are singular, i.e. in every gauge there 
is a singularity of the Yang--Mills curvature where the monopole is 
located. The position of the singularity is then independent of the choice 
of the $U(1)$ subgroup, or equivalently of the abelian gauge projection, 
used to define the magnetic charge and it identifies the position of a 
${\bf Z}_2$--monopole.

\quad

The definition of magnetic charge and ${\bf Z}_2$-- 
charge can be easily adapted to the lattice formulation as follows [19].

Let $g$ denote the $SU(2)$ lattice gauge field. The Yang--Mills action on 
the lattice is given by

$$
S_{YM} (g) = \beta \sum_p \bigl(1-\chi (g_{\partial p})\bigr) \eqno(4.3)
$$

where $\chi$ is the character of the fundamental representation.

\quad
Denote by $X (g)$ the scalar field with values in $su(2)$ transforming 
under the adjoint representation, identifying the abelian projection gauge.
Let $W$ be the $SU(2)$ gauge transformation such that for every site $i$ on
the lattice one has 

$$
W_i X_i (g) W^+_i = |X_i (g)| \sigma_3,\eqno(4.4)
$$

were $|X_i| = \sqrt{Tr X^a_i X^a_i}$ and define

$$
\tilde g_{<ij>} = W_i g_{<ij>} W^+_j. \eqno(4.5)
$$

$\tilde g$ denotes the $SU(2)$--gauge field in the abelian projection
gauge defined by $X$.

$\tilde g_{<ij>}$ can be decomposed as a product of two matrices $C_{<ij>}$
and $u_{<ij>} (\theta)$, where 

$$
C_{<ij>} =\left(\matrix{(1-|c_{<ij>}|^2)^{1\over2} & -\bar c_{<ij>}\cr
c_{<ij>}& (1- |c_{<ij>}|^2)^{1\over2}\cr}\right),\quad
u_{<ij>}= e^{i {1 \over 2} \theta_{<ij>} \sigma_3}, \eqno(4.6)
$$

with $c_{<ij>} \in {\bf C}, \bar c_{<ij>}$ denoting its complex 
conjugate, and ${1 \over 2} \theta_{<ij>}= \arg (\tilde g_{<ij>})_{11}$.

\quad
Hence $\theta$ is a $U(1)$ gauge field with range $(-2\pi, 2\pi)$ and $c$ 
is a charged field of charge 1.

A similar decomposition holds for the coset variable given on a link $<ij>$
by $g_{<ij>} \Gamma \equiv U_{<ij>}$, where $\Gamma$ is the centre of 
$SU(2)$, isomorphic to ${\bf Z}_2$. $U_{<ij>}$ can be viewed as an 
$SO(3)$--gauge field and in the decomposition (4.6) $\theta$ is now a 
$U(1)$ field with range $(-\pi, \pi)$.

We define the magnetic charge in a lattice cube $c$ by

$$
m_c (\theta) = {1\over 2\pi} \sum_{p\in\partial c} (d\theta)_p
\eqno(4.7)
$$

where $(d\theta)_p$ is restricted to the range $(-\pi,\pi)$.

The ${\bf Z}_2$--charge in $c$ is defined by 

$$
e^{i\pi z_c(g)} = e^{i\sum_{p\in\partial c} \arg \chi(g_{\partial p})}. 
\eqno(4.8)
$$

A plaquette $p$ where 

$$
e^{i\arg \chi (g_{\partial p})} = -1 \eqno(4.9)
$$

can be identified as the location of a Dirac string of a ${\bf Z}_2$-- 
monopole intersecting the plane containing $p$.

\quad
In the lattice formulation the relation discussed in the continuum between
${\bf Z}_2$-- and magnetic charge becomes

$$
e^{i\pi z_c (g)} = e^{i\pi m_c (\theta)} \eqno(4.10)
$$

\vskip 0.3truecm

{\bf 5. The $U(1)$--monopole order parameter for $SU(2)$ Yang--Mills theory}

\vskip 0.3truecm

\quad
In this section we recall some attempts made to derive an order parameter 
in $SU(2)$ Yang--Mills theory on the lattice, using a construction
directly inspired by Dirac ansatz.

\quad
Integrating out the charged field $c$ defined in (4.6) one can view the 
$SU(2)$ gauge theory in an abelian projection gauge as a $U(1)$ theory with
gauge field $\theta$ and an effective action of the form 
$S_{eff}(d\theta)$, where we used the residual $U(1)$ gauge invariance to 
deduce the dependence on the curvature $d\theta$.

\quad
It is then natural, following the dual of Dirac ansatz, to try to define 
the two--point Green function of the magnetic monopole of the abelian 
projection by performing the shift 

$$
d\theta \rightarrow d\theta + 4 \pi \delta \Delta^{-1} (B^x - B^y + \omega) 
\eqno(5.1)
$$

where $\omega$ is a 3--form Poincar\'e dual of a path connecting $x$
to $y$.

\quad
This is basically the attempt made in [6] in the Maximal
Abelian Gauge [19] and in [10] in the Spatial Maximal Abelian Gauge.

However, a closer look shows some inconsistency in this approach.
In fact, it is easy to prove that the effective action is given by an
expression of the form
$$
\sum_{\cal L} C_{\cal L} e^{i {1 \over 2} \sigma_3 \sum_{<ij>\in{\cal L}}
\theta_{<ij>}} \eqno(5.2)
$$

where ${\cal L}$ is a loop and $C_{\cal L}$ a complex coefficient 
independent of $\theta$.

Choosing, for each ${\cal L}$, a surface $\Sigma ({\cal L)}$ such that 
$\partial\Sigma({\cal L)} = {\cal L}$, one can rewrite (5.2) as

$$
\sum_{\cal L} C_{\cal L} e^{i{1 \over 2} \sigma_3 \sum_{p\in\Sigma({\cal 
L})} (d\theta)_p},
$$

i.e. in the form $S_{eff} (d\theta)$ and the choice of the surfaces
$\Sigma ({\cal L})$ is obviously irrelevant.
 
\quad
However, when the shift (5.1) is performed the term

$$
e^{i \sigma_3 \sum_{p\in\Sigma({\cal L})} 2 \pi \delta\Delta^{-1} 
(B^x-B^y+\omega)} \eqno(5.3)
$$

is no more independent of the choice of $\Sigma({\cal L})$. This 
inconsistency is due to the violation of Dirac quantization condition at 
the level of currents related to $B^x - B^y $.

\quad
The approach of [7,9] is slightly different and monopole Green functions
are defined directly by modifying the Yang--Mills action.

Let us fix a convention assigning to every plaquette $p$ a site $j(p)$ on 
its boundary (see [9] for more details) and let $X (g)$ denote the scalar 
field defining the abelian projection.

\quad
Within our setting the proposal made in [7,9] to obtain the
monopole--monopole Green 
function is to replace the plaquette term $\chi (g_{\partial p})$ in the 
Yang--Mills action (4.3) by

$$
\chi \Bigl(g_{\partial p} e^{i ({X\over |X|})_{j (p)} 
2 \pi [\delta \Delta^{-1} (B^x - B^y + \omega)]_p}\Bigr). 
\eqno(5.4)
$$

Actually, in [7,9] the support of $\omega$ consists of a sum of the
dual of two straight lines, $\omega^x$ and $-\omega^y$ each at constant
time in 
the 3--direction. With this choice $B^x+\omega^x$ can be related, as in 
(2.18), to the magnetic field of a Dirac monopole with its Dirac string 
along the 3--direction.

\quad
In the abelian projection gauge defined by $X$, using the decomposition 
(4.6), one can rewrite the argument of $\chi$ in (5.4) as

$$
\prod_{<ij> \in\partial p} \bigl(e^{i \vec c_{<ij>} \cdot  
\vec\sigma}
e^{i {1 \over 2}
\theta_{<ij>} \sigma_3}\bigr) e^{i \sigma_3 2 \pi [\delta
\Delta^{-1} 
(B^x-B^y + \omega)]_p} ,
\eqno(5.5)
$$

where

$$
c_{<ij>}= c^1_{<ij>}+ i c^2_{<ij>}, \quad c^\alpha_{<ij>},  \in {\bf R}, 
\alpha=1,2 \quad 
\vec c = (c^1, c^2). 
$$
We insert the lattice constant $\epsilon$ and, since we are really
interested
in the continuum limit, we consider an expansion in $\epsilon$ up to
$0(\epsilon^2)$:
One finds

$$
({\rm 5.5}) \sim \prod_{{<ij>}\in\partial p} \bigl(e^{i\vec c_{<i j>}
\cdot \vec\sigma \epsilon}\bigr)\,
e^{i \sigma_3 2 \pi [d\theta + \delta\Delta^{-1} 
(B^x-B^y+\omega)]_p \epsilon} \bigl(1+O(\epsilon^2)\bigr).
$$

\quad
Therefore to $O(\epsilon^2)$ this recipe seems to give a consistent 
prescription; however problems arise at order $\epsilon^2$ because in 
$g_{\partial p}$ there appear terms depending on $\theta$ which cannot be 
rewritten in terms of $(d\theta)_p$:

$$
O(\epsilon^2)= \epsilon^2 \{[(d\theta)_p \sigma_3, \vec\gamma (c) \cdot 
\vec\sigma]+ (\theta\sigma_3 \wedge \vec\gamma{(c)} \cdot \vec\sigma)_p\}
+ O(\epsilon^3), \eqno(5.6)
$$

where $\wedge$ is the wedge product on the lattice (see e.g. [10]) and 
$\vec\gamma = (\gamma^1, \gamma^2)$ are functions of $c$.

As a consequence of the last term in (5.6), a change in the choice of the 
``Dirac string" $\omega$ cannot be eliminated by a shift in $\theta$, 
and again this can be traced back to a violation of Dirac quantization 
condition at the level of currents.

\quad
The consistency to order $\epsilon^2$ of the recipe in [7,9] suggests
however
that this disorder parameter (and the one of [6]) could be meaningful at
large scales, as 
supported by numerical results for the (physical) temperature of the 
deconfinement transition.

\vskip 0.3truecm
{\bf 6. Green functions for regular monopoles  in SU(2) lattice gauge 
theory}

\vskip 0.3truecm

\quad
Combining the ideas of the last three sections, one is lead to propose a 
definition of Green functions for regular monopoles of any  abelian 
projection, as follows.

\quad
Imagine that a regular charge--2 monopole in an abelian projection gauge is
created at a site $x$ and annihilated at a site $y$.
We propose to construct the corresponding two--point function summing over 
a pair of fluctuating strings carrying magnetic flux 1 with end points at 
the position of the monopole.

\quad
As discussed in section 4, strings of odd magnetic flux in every abelian 
projection can be identified as Dirac strings of ${\bf Z}_2$--monopoles, 
and hence are independent
 of the choice of the projection.

\quad
In turn, ${\bf Z}_2$--monopoles can be introduced by means of a 't Hooft 
disorder fields which, in $SU(2)$ lattice gauge theory, is defined as 
follows.

\quad
Let $\Sigma$ be the Poincar\'e  dual of a surface bounded by a loop ${\cal 
L}$. Then the corresponding disorder field is defined by

$$
D(\Sigma) = e^{-[S_{YM} (\{g_{\partial p} e^{i\pi\Sigma_p\sigma_3}\}) -
S_{YM} 
(\{g_{\partial p}\})]} \eqno(6.1)
$$

where $S_{YM}$ is the action (4.3).
The expectation value of $D(\Sigma)$ depends only on ${\cal L}$ and 
describes the worldlines of a ${\bf Z}_2$ monopole--antimonopole pair.

\quad
Since the regular monopoles have even magnetic charge and the charged 
field, $c$, of the abelian projection has integer  electric  charge,
one can
adapt to the present setting the construction presented in section 3 for 
$q=2$.

\quad
Let $j_{xy}$ be an integer 1--current satisfying $\delta j_{xy} = \delta_y 
- \delta_x$ with support on a path connecting $x$ to $y$, and let ${\cal
D}
\nu_E (\gamma^x)$, ${\cal D}\nu_E (\gamma^y)$ denote normalized, signed measures 
over ${\bf Z}/2$--valued currents $\gamma^x, \gamma^y$, constrained by
$\delta \gamma^z = \delta z, z=x,y$
as defined in section 3.

\quad
To a configuration $\{\gamma^x, \gamma^y, j_{xy}\}$ we associate a
2--current $\Sigma
(\gamma^x-\gamma^y+j_{xy})$ satisfying 

$$
^* d\Sigma (\gamma^x-\gamma^y+j_{xy}) = 2(\gamma^x-\gamma^y+j_{xy}),
$$
where the 2 in the.h.s. appears for agreement with definition (6.1).
$\Sigma$ is then the Poincar\'e  dual
of a two--sheet surface, with boundary given by the support of $\gamma^x - 
\gamma^y$ and the two sheets joining each other along the support of 
$j_{xy}$.

\quad
We define the monopole two--point correlation function by 

$$
G(x,y) = \int {\cal D} \nu_E (\gamma^x) {\cal D}\nu_E (\gamma^y)
\langle D(\Sigma(\gamma^x-\gamma^y+ j_{xy})) \rangle.\eqno(6.2)
$$

We claim that long--range order in this correlation functions 
characterizes the confinement phase.

\quad

From the $n$--point monopole correlation functions defined by
generalizing 
eq. (6.2) one can (at least formally) reconstruct, via O.S. lattice 
reconstruction, a monopole field operator $\hat m(B^x), B=^*E$. The long--range 
order for $G(x,y)$ then corresponds to a non--vanishing vacuum expectation
value of $\hat m$.

\quad
Since equation (6.2) does not involve any abelian projection, it 
follows that, while the definition of the trajectory of a regular monopole 
requires choosing an abelian projection, the locations of creation and 
annihilation of a monopole are independent of that choice, hence intrinsic 
to the $SU(2)$ theory.

\quad

From numerical simulations one may gain some indirect support for the 
conjecture that $\hat m$ is a good order parameter for the 
confinement--deconfinement transition by 
noticing that the large distance behaviour of $G(x,y)$ appears to approach 
that of the $U(1)$--monopole Green functions of [7].

\quad
In fact, since the group manifold of $SU(2)$ is isomorphic, via the 
exponential map, to a 3--ball of radius $\pi$ with boundary points
identified, one may 
replace  $e^{i\pi\Sigma_p \sigma_3}$ by $e^{i\pi ({X\over |X|})_{ j(p)} 
\Sigma_p}$
in the definition (6.2) of the disorder field for any choice of scalar $X$ 
defining an abelian projection. This is because $({X\over |X|})_{j(p)}$ 
defines a unit vector in $su(2)$.

\quad
As remarked in section 3, at large scales the measure ${\cal D} \nu_E
(\gamma^x) $  behaves as an approximate Dirac measure peaked around
the current configuration $E^x$, and this, in turn, implies that in 
(6.2) the configurations of $d\Sigma (\gamma^x-\gamma^y+j_{xy})$ are
peaked around $2 (B^x-B^y +\omega)$, with $B=^*E, 
\omega=^* j_{xy}$.

Therefore, in the scaling limit, one expects that

$$
\int{\cal D} \nu_E(\gamma^x) {\cal D} \nu_E (\gamma^y) e^{-S_{YM}
[\{g_{\partial p} 
e^{i\pi ({X\over |X|})_{j(p)} \Sigma_p}\}]}
$$
$$
\sim e^{-S_{YM}[\{g_{\partial p} e^{i ({X\over |X|})_{j(p)} 
2 \pi [\delta\Delta^{-1} (B^x - B^y +\omega)]_p}\}]}.\eqno(6.3)
$$

This would reproduce the behaviour of the disorder field defined via eq.
(5.4). 
Numerically, its expectation value provides a clear signal of the 
confinement--deconfinement transition [7]. Eq. (6.3) yields an explanation
of the numerical evidence that, from order parameters corresponding 
to different abelian projections, one obtains the same transition temperature
[9].

\quad
In fact, this transition is governed by the low--energy physics, correctly 
captured by the scaling limit. Hence, assuming that eq. (6.3) holds, the
order 
parameters defined via eq. (5.4), unphysically dependent on the choice of 
the ``Dirac string $\omega$" at small scales, are just the scaling 
limit of the order parameter defined via (6.2), which is manifestly 
independent of the choice of an abelian projection and of the ``Dirac 
string $j_{xy}"$.

\quad
Expressing $\langle D \Bigl(\Sigma
(\gamma^x-\gamma^y+j_{xy})\Bigr)\rangle$ in terms 
of magnetic currents, through a duality transformation, one can exhibit 
more explicitly the non--vanishing vacuum expectation value of $\hat m$ as
a dual 
Higgs mechanism (in the spirit of the ``abelian dominance" scenario).

\quad
Since, however, $\Sigma (\gamma^x-\gamma^y+j_{xy})$ can be interpreted as
a (double)
sheet of center vortices, a connection with the ``center dominance" 
scenario emerges, as discussed in next section.

\quad
To derive the duality transformation, we express the correlation function 
$\langle D(\Sigma)\rangle$, where $\Sigma \equiv \Sigma 
(\gamma^x-\gamma^y+j_{xy}$), in terms of the variable $\theta$ and $c$ 
defined in eq. (4.6), and insert in the integration measure 
an abelian projection gauge fixing

$$
\delta ({X\over |X|} (c,\theta) - \sigma_3).
$$

First we integrate out $c$. As a consequence of the residual $U(1)$ gauge 
invariance we can expand

$$
\int \prod_{<ij>} dc_{<ij>} d \bar c_{<ij>} e^{-S_{YM} (\{g_{\partial p}
(\theta,c) e^{i\pi\Sigma_p\sigma_3}\})}
\prod_j \delta \Bigl({X_j\over|X_j|} (\theta,c)-\sigma_3\Bigr)\eqno(6.4)
$$

as a Fourier series in $d\theta+2\pi\Sigma$.

The coefficients are denoted by $F(n)$, where $n$ is a ${\bf Z}/2$--valued 
2--form.

We define a 1--form $\ell$ by

$$
\delta n=\ell, \eqno(6.5)
$$

and decompose the 2--form $n$ as

$$
n=n[\ell] + ^*d\xi, \eqno(6.6)
$$

where $n[\ell]$ is a ${\bf Z}/2$--valued solution of (6.5) and $\xi$ is a 
${\bf Z}/2$--valued 1--form in the dual lattice. Then one obtains

$$
\langle D (\Sigma) \rangle ={1\over Z} \sum_{[\xi]} 
\sum_{\ell:\delta\ell=0} F(n[\ell]+ ^*d\xi)
\int \prod_{<ij>} d\theta_{<ij>} e^{i(\ell,\theta)} e^{i 2 \pi (\Sigma,
^*d\xi)}, 
\eqno(6.7)
$$

where $[\xi]$ denotes a gauge equivalence class of $\xi$, and the equation

$$
(n[\ell], d\theta)=(\ell,\theta)
$$

has been used.

\quad
Integrating over $\theta$ imposes the constraint $\ell=0$. Hence, in 
particular, one can choose $n[\ell]=0$. Furthermore we can replace $\xi$ by
a real--valued 1--form $A$ by inserting the term $\sum_{\rho:\delta\rho=0} 
e^{i4\pi(A,\rho)}$, where $\rho$ is a {\bf Z}--valued 1--form (see e.g. 
[21]).

\quad
As a result we obtain

$$
\langle D\Bigl(\Sigma(\gamma^x-\gamma^y+j^{xy})\Bigr)\rangle=
{1\over Z} \int d[A] \sum_{\rho:\delta\rho=0} F(^*dA)
e^{i 4\pi (\gamma^x-\gamma^y+j_{xy}+\rho,A)},
\eqno(6.8)
$$

where $d[A]$ denotes formal integration over gauge equivalence classes of 
$A$.

In (6.8), worldlines of regular monopoles of the abelian projection are 
described by the currents $j_{xy}+\rho$, and they exhibit sources at $\{x\}$ 
and $\{y\}$.

The representation (6.8) explicitly proves independence of the choice of 
$j_{xy}$ in the construction of Green functions.

By setting

$$
\tilde S (A) \equiv -\ln F(^*dA)
$$

one can view the dual model appearing in (6.8) as a Higgs model with gauge 
action $\tilde S(A)$ and the correlation function

$$
\int {\cal D} \nu_E (\gamma^x){\cal D} \nu_E (\gamma^y) \langle D(\Sigma 
(\gamma^x - \gamma^y+j_{xy}))\rangle
$$

can be viewed as the two--point function of the charged field, 
$\langle\phi(E^x) \bar\phi (E^y)\rangle^\sim$, of that dual model, where 
$\langle \cdot \rangle^\sim$ denotes the corresponding expectation value.

\quad
The abelian Higgs model in four dimensions has two phase, the
Coulomb 
and the Higgs phase. If the dual model is in the Higgs phase one expects
that
$\langle \phi (E^x)\rangle^\sim \not=0$.
(In fact, for the standard gauge action and with the Dirac recipe for the 
charged field, this has been proved in [13], [16].) In the original model 
this non--vanishing expectation value corresponds to $\langle\Omega| \hat 
m (B^x) \Omega \rangle\not =0$, thus describing monopole condensation. 

\quad
The suggestion that the dual model is in the Higgs phase comes  
from numerical simulations, as previously discussed.

The above construction makes the relation between the original $SU(2)$ 
gauge theory and a dual abelian Higgs model more precise, as advocated by 
many authors.

\quad
We end this section with a remark about about monopole Green functions in 
the formal continuum limit.
Our discussion of monopole in $SU(2)$--Yang Mills theory was performed in 
the lattice, because it heavily relies upon the 't Hooft disorder field 
which has no simple continuum analog.

\quad
Presumably one can construct  a disorder field with the desired properties 
in the continuum using the loop space formalism developed in [22], 
but this will be 
discussed elsewhere.

\vskip 0.5truecm
{\bf 7. Relation with center dominance}
\vskip 0.3truecm

\quad
The representation (6.8) does not exhibit center vortex  sheets; they are 
hidden in the definition of $F(^*dA)$. To exhibit them explicitly one 
starts, following [23], by replacing the $SU(2)$--gauge field $g$ with a 
couple of new variables $\{U,\sigma\}$ where $U$ is the gauge coset 
variable, introduced in sect.4, which can be viewed as an $SO(3)$ gauge 
field and $\sigma$ is a 2--form with values in $\{0,1\} \simeq {\bf Z}_2$.

\quad
It has been shown in [22] that the two fields $U$ and $\sigma$ are not 
independent. It is easy to show that $\sum_{p\in\partial c} {\rm arg} 
\chi(g_{\partial p})$ is only a function of the coset field $U$, which we
denote 
by $z_c(U)$ and the following constraints holds:

$$
e^{i\pi(d\sigma)_c} = e^{iz_c(U)} \eqno(7.1)
$$

for each cube $c$.

Let us discuss the relation between $\sigma$ and the 't Hooft disorder 
field.

The plaquette term $\chi(g_{\partial p} e^{i\pi\sigma_3\Sigma_p})$ 
appearing in $\langle D(\Sigma)\rangle$ is rewritten in the new variables, 
as

$$
|\chi| (U_{\partial p}) e^{i\pi(\sigma_p+\Sigma_p)}, \eqno(7.2)
$$

where $|\chi|(U)= |\chi(g)|.$

Hence the introduction of the disorder field $D(\Sigma)$ induces a shift of
$\sigma_p$ by $\Sigma_p$.

\quad
The constraint (7.1) can be solved by 

$$
e^{i\pi\sigma_p} = {\rm sign} \chi (U_{\partial p}) \prod_{<ij>\in\partial
p} {\rm sign} 
\chi (U_{<ij>})
$$
$$
= {\rm sign}  \chi (g_{\partial p}) \prod_{<ij>\in\partial p} {\rm sign}
\chi 
(g_{<ij>})\eqno(7.3)
$$

This solution is gauge--dependent but does not involve any choice of 
abelian projection.

\quad
In the center dominance scenario one defines the maximal center gauge in 
$SU(2)$ gauge theory as the gauge which brings the link variables
$\{g_{<ij>}\}$
as close as possible to the center, $\Gamma \simeq {\bf Z}_2$, of $SU(2)$, 
by maximizing the quantity

$$
R= \sum_{<ij>} Tr (g_{<ij>})^2.
$$

In the maximal center gauge, a plaquette $p$ where 
$$
\prod_{<ij>} {\rm
sign} \chi
(g_{<ij>})=-1
\eqno(7.4)
$$
is the location of a P--vortex.

It has been rigorously proven in [23] that, for large $\beta$, i.e., close 
to the continuum limit, the set of plaquettes where sign $\chi(g_{\partial
p})=-1$ is dilute.

\quad
Therefore, since sign $\chi(g_{\partial p})$ is gauge--invariant, for large
$\beta$, the identification of a P--vortex location as the set of 
plaquettes where $\sigma_p\not= 0$ in the center projection gauge, should 
be equivalent to the standard definition, equation (7.4), from the point
of view of 
discussing  the deconfinement transition.

\quad
Numerical simulations shows that, in the confining phase, P--vortex sheets 
percolate [2]. This suggests that in this phase even the introduction of an
additional infinite P--vortex sheet $\Sigma$, like the one involved in the 
construction of the monopole Green function, should be a small perturbation
and should not lead to a clustering behaviour.

\quad
In other words one could interpret a non--vanishing expectation value of 
the monopole operator $\hat m$ in the maximal center gauge as due to a 
condensation of P--vortex sheets, in the spirit of center dominance.

In the deconfinement phase at positive temperature, numerical simulations 
shows that P--vortex sheets are dilute [2]. Hence, for large $\beta$, it is
natural to conjecture that the introduction of an infinite P--vortex sheet 
$\Sigma$ leads to clustering and, as a consequence, to a vanishing 
expectation value for $\hat m$.

\quad
Finally we remark that an approximate relation between P--vortex sheets and
regular monopole worldlines can be established following [24].

The double--sheet P--vortex structure associated to monopole worldlines 
appearing in [24
] is a natural counterpart of our construction of
monopoles
in terms of the double--sheet surface $\Sigma$.

\vfill\eject

{\bf Appendix}

\vskip 0.5truecm
\underbar{Forms and currents in the continuum and on the lattice}

\vskip 0.3truecm

In the continuum we consider as `` space--time" the euclidean space ${\bf 
R}^d$. 

Given an antisymmetric tensor field of rank $k$ on ${\bf R}^d, 
a_{\mu_1... \mu_k} (x)$ one defines the associated $k$--form by setting

$$
a^{(k)} (x) = {1\over k!} a_{\mu_1 ...\mu_k} (x) dx^{\mu_1} \wedge...
\wedge 
dx^{\mu_k} \eqno(A.1)
$$

where $\wedge$ is the wedge (antisymmetric tensor) product. The space of 
$k$--forms is a group $\Lambda^k({\bf R}^d)$ under pointwise addition. 
We denote by

$$
d : \Lambda^k ({\bf R}^d) \rightarrow \Lambda^{k+1} ({\bf R}^d)
$$

the exterior differential defined through

$$
da^{(k)} (x) = {1\over(k+1)!} \partial_\mu a_{\mu_1 ... \mu_k} (x) dx^\mu 
\wedge dx^{\mu_1}\wedge...
dx^{\mu_k}, \eqno(A.2)
$$

by

$$
* : \Lambda^k ({\bf R}^d) \rightarrow \Lambda^{d-k} ({\bf R}^d)
$$

the Hodge star defined through

$$
^*(a^{(k)} (x)) = {1\over k!} {1\over (d-k)!} 
\epsilon_{\mu_1...\mu_{d-k+1}...\mu_d} 
a^{\mu_{d-k+1}...\mu_d} (x) dx^{\mu_1} \wedge \wedge dx^{\mu_{d-k}},
\eqno(A.3)
$$

by $\delta =^*d^* (-1)^{d(k+1)}$ the codifferential and by

$$
\Delta = \delta d+ d\delta \eqno(A.4)
$$
the Laplacian.

An inner product between $k$--forms is defined by setting

$$
(a^{(k)}, b^{(k)}) = \int d^d x \, a_{\mu_1...\mu_k} (x) b^{\mu_1...\mu_k}
(x)
= \int a^{(k)} \wedge^* b^{(k)} \eqno(A.5)
$$

and it satisfies

$$
(a^{(k)}, db^{(k-1)}) = (\delta a^{(k)}, b^{(k-1)}). \eqno(A.6)
$$

The $L^2$ norm corresponding to the inner product (A.5) 
is denoted by $|| \ ||$, i.e. 
$(a^{(k)}, a^{(k)}) \equiv || a^{(k)} ||^2$.

\quad
The Poincar\'e  lemma states that if  $da^{(k)} =0$, then there exist 
$a^{(k-1)}$ such that $a^{(k)} = da^{(k-1)}$.

A $k$--current in ${\bf R}^d$ is a linear functional in the space of 
$d-k$ 
forms with compact support, continuous in the sense of distributions, i.e. 
$k$--current are $k$--forms with distribution--valued components [25].

\quad
In the space of currents there exist a map, Poincar\'e  duality,  
associating to a $k$--dimensional surface $\Sigma_k$ a
$(d-k)$--current, 
$PD({\Sigma_k})$, according to 

$$
\int_{\Sigma_k} a^{(k)} = \int_{{\bf R}^d} a^{(k)} \wedge PD({\Sigma_k}),
\eqno(A.7)
$$

for any $k-$form $a^{(k)}$ of compact support. The following property 
holds:

$$
PD({\partial\Sigma_k}) = d PD({\Sigma_k}) \eqno(A.8)
$$

where $\partial$ denotes the boundary operator.

A basic consequence of Poincar\'e's  duality is that, whenever well
defined,

$$
\int_{{\bf R}^d} PD({\Sigma_k}) \wedge PD({\Sigma_{d-k}})
$$

is an integer counting the intersection with sign of $\Sigma_k$ with 
$\Sigma_{d-k}$. 

Linear combinations of such $k$--currents $PD({\Sigma_k})$ with integer 
coefficients are called integral $k-$currents.
Poincar\`e's lemma holds also for currents.

\quad
We now turn to the lattice. 

\quad
Our lattice is ${\bf Z}^d_{1/2}$, where the subscript 1/2 indicate that
the
coordinates of the sites are half--integer.

If $W$ is an additive abelian group, one can define $k$--forms with values 
in $W$ as maps, $a^{(k)}$, from oriented $k$--cells, $c_k$, of the
lattice 
to $W$ satisfying $a^{(k)} (-c_k) = -a^{(k)} (c_k)$, where $-c_k$ denotes
the cell obtained from $c_k$ reversing the orientation.

We denote by $d$ the lattice exterior differential:

$$
d a^{(k)} (c_{k+1})= \sum_{c_k\in\partial c_k+1} a^{(k)} (c_k) \eqno(A.9)
$$

and by ${}^*$ the Hodge star. Let $c^*_{d-k}$ denote the cell in the dual 
lattice, ${\bf Z}^4$, dual to $c_k$. Then

$$
^*(a^{(k)}) (c^*_{d-k})= a^{(k)} (c_k). \eqno(A.10)
$$

We also introduce the (lattice) codifferential $\delta = (-)^{d(k+1)}
{}^*d^*$
and 
Laplacian $\Delta = d\delta + \delta d$.

If $W$ is a Hilbert space with inner product ( , ) one can define a inner 
product among $W$--valued $k$--forms $a^{(k)}$ and $b^{(k)}$ by

$$
(a^{(k)}, b^{(k)}) = \sum_{c_k} (a^{(k)}, b^{(k)}). \eqno(A.11)
$$

The $\ell_2$--norm corresponding to this scalar product is denoted by  $|| 
\ ||$. Equation $(A.6)$ and the Poincar\'e  lemma hold also on the lattice.

If $W$ is a discrete group we call the $W$--valued $k-$forms also 
$k-$currents, in analogy with the continuum definition.

If $\Sigma_k$ is a $k-$dimensional surface in the lattice one defines its 
Poincar\'e  dual as the $d-k$ current $PD({\Sigma_k})$ in the dual
lattice
such that

$$
PD({\Sigma_k}) (c^*_k)= \cases{1 & if $c_k \in \Sigma_k$\cr
0 & otherwise. \cr} \eqno(A.12)
$$

\quad
In the paper we do not make use of the symbol $PD$, introduced in this
appendix for sake of clarity, and often identify a $k$--surface with its
Poincar\'e dual defined above.

\vskip 1truecm
{\bf Acknowledgments}

Useful discussions with M. Chernodub, Ph. de Forcrand, A. Di Giacomo, M.
Polikarpov and M. Vettorazzo are gratefully acknowledged. One of us (P.M.)
thanks the E.T.H. in Z\"urich for kind hospitality.
This work was partially supported by the European Commission RTN programme
HPRN-CT2000-00131. to which P.M.is associated
 
\vskip 1truecm
{\bf References}
\vskip 0.5truecm
\item{[1]} G. 't Hooft, Nucl. Phys. B \underbar{153}, 141 (1979);

G. Mack in ``Recent Developments in Gauge Theories", (Carg\'ese 1979), 't 
Hooft ed. Plenum Press 1980; 

A.M. Polyakov, Phys. Lett. \underbar{72B}, 477 (1978)

H. Nielsen, P. Olesen, Nucl. Phys. \underbar{B160}, 360 (1979).

\item{[2]} L. Del Debbio, M. Faber, J. Greensite, S. Oleynik, Phys. Rev. 
\underbar{D55}, 2298 (1997).

\item{[3]} D. Durhuus, J. Fr\"ohlich, Commun. Math. Phys., \underbar{75}, 
103 (1983).

\item{[4]} S. Mandelstam, Phys. Rep. \underbar{23C}, 245 (1976).

\item{[5]} G. 't Hooft, Nucl. Phys. \underbar{B190}, 455 81981)

\item{[6]} M. Chernodub, M. Polikarpov, M. Zubkov, Nucl. Phys. Proc. 
Suppl. \underbar{34}, 256 (1994);

\item{[7]} L. Del Debbio, A. Di Giacomo, G. Paffuti, P. Pieri, Phys. Lett. 
\underbar{B355}, 255 (1995); G. Di Cecio, A. Di Giacomo, G. Paffuti, M. 
Trigiante, Nucl. Phys. \underbar{B489}, 739 (1997);
M. Chernodub, M. Polikarpov, A. Veselov, Phys. Lett. B \underbar{399}, 267 
(1997).

\item{[8]} P.A.M. Dirac, Can. J. Phys. \underbar{33}, 650 (1955).

\item{[9]} A. Di Giacomo, B. Lucini, L. Montesi, G. Paffuti, Nucl. Phys. 
Proc. Suppl. \underbar{63}, 540 (1998).

\item{[10]} J. Fr\"ohlich, P.A. Marchetti, Nucl. Phys. \underbar{B511}, 770 
(1999).

\item{[11]} G. Morchio, F. Strocchi, Nucl. Phys. \underbar{B211}, 471 
(1983).

\item{[12]} K. Osterwalder, R. Schrader, Commun Math. Phys. 
\underbar{31}(1973); \underbar{42}, 281 (1975).

E. Seiler ``Gauge Theories as a Problem in Constructive Quantum Field 
Theory and Statistical Mechanics". Lecture Notes in Physics 159, Springer 
Verlag 1982.

\item{[13]} J. Fr\"ohlich, P.A. Marchetti, Europhys. Lett. \underbar{2}, 
933 (1986); in ``Algebraic Theory of Superselection Sectors. Introduction 
and Recent results". ed. D. Kastler, World Scientific 1990.

\item{[14]} F. Wegner, J. Math. Phys. \underbar{12}, 2254 (1971);

G. 't Hooft, Nucl. Phys. \underbar{B138}, 1 (1978).

\item{[15]} J. Fr\"ohlich, P.A. Marchetti, Commun. Math. Phys. 
\underbar{112}, 343 (1987).

\item{[16]} T. Kennedy, C. King, Phys. Rev. Lett. \underbar{55}, 776
(1985)

\item{[17]} S. Mandelstam, Ann. Phys. \underbar{19}, 1(1962).

\item{[18]} S. Coleman in ``The Unity of the Fundamental Interactions" 
(Erice 1981), A. Zichichi ed. Plenum 1983.

\item{[19]} A.S. Kronfeld, G. Schierholz, U.J. Wiese, Nucl. Phys. 
\underbar{B292}, 461 (1987).

\item{[20]} M. Chernodub, M. Polikarpov, A. Veselov, JETP \underbar{69}, 
174 (1999).

\item{[21]} J. Fr\"ohlich, T. Spencer, Commun. Math. Phys. \underbar{83}, 
411 (1982).

\item{[22]} Chen Hong--Mo, J. Faridani, Tsou Sheung Tsun, Phys.  Rev. 
\underbar{D53}, 7293 (1996).

\item{[23]} G. Mack, V. Petkova, Z. Phys. C \underbar{12}, 177 (1982)

\item{[24]} M. Engelhardt, H. Reinhardt, Nucl. Phys. \underbar{B567}, 249 
(2000);

C. Alexandrou, M. D'Elia, Ph. de Forcrand, Nucl. Phys. Proc. Suppl. 
\underbar{83}, 437 (2000).

\item{[25]} G. de Rham ``Differential Manifolds. Forms, Currents, 
Harmonic Forms", Springer Verlag 1984.

\bye